\documentclass[a4paper,conference]{IEEEtran}
\usepackage{stfloats}
\usepackage{booktabs}
\usepackage[ruled]{algorithm2e} 
\usepackage{graphicx}
\usepackage{caption}
\usepackage{subcaption}
\usepackage{float}
\usepackage{lipsum} 
\renewcommand\IEEEkeywordsname{Keywords}

\ifCLASSINFOpdf
\else
\fi
\usepackage{booktabs}

\usepackage{booktabs}
\usepackage{array}
\begin{document}
%
\title{DeepHS-HDRVideo: Deep High Speed High Dynamic Range Video Reconstruction}


\newcommand\BibTeX{B\textsc{ib}\TeX}


%


\author{\IEEEauthorblockN{Zeeshan Khan\IEEEauthorrefmark{1},
Parth Shettiwar\IEEEauthorrefmark{2}, Mukul Khanna\IEEEauthorrefmark{3} and
Shanmuganathan Raman
\IEEEauthorrefmark{4}}
\IEEEauthorblockA{\IEEEauthorrefmark{1}CVIT, IIIT Hyderabad,
\IEEEauthorrefmark{2}University of California, Los Angeles\\
\IEEEauthorrefmark{3}Georgia Tech,
\IEEEauthorrefmark{4}IIT Gandhinagar\\
zeeshan.khan@research.iiit.ac.in,
parthshettiwar@g.ucla.edu,\\
mukul18khanna@gmail.com,
shanmuga@iitgn.ac.in}}


\let\oldtwocolumn\twocolumn
\renewcommand\twocolumn[1][]{%
    \oldtwocolumn[{#1}{
    \begin{center}
           \includegraphics[width=6.8in]{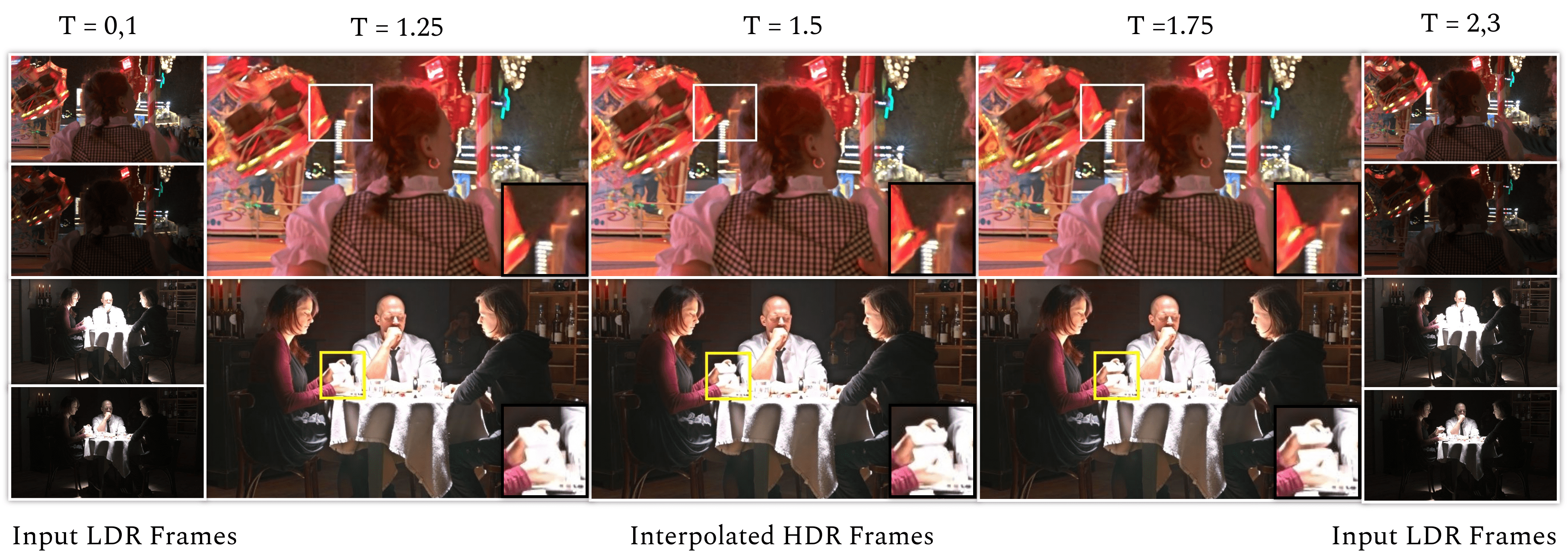}
            \captionof{figure}{Example of  high speed HDR video generation using video interpolation. From a sequence of alternating exposure LDR frames at $T = 0,1,2,3$, we generate multiple intermediate HDR frames between $T = 1$ and $T = 2$. Video frame interpolation allows us to augment the dynamic range at each timestamp, and recursively upscale the video frame-rate manifold. Our method can generate high quality HDR videos at arbitrarily high frame rates. More results are presented in the supplementary video.}

           \label{fig:fig1}
        \end{center}
    }]
}



\maketitle
\begin{abstract}
Due to hardware constraints, standard off-the-shelf digital cameras suffers from low dynamic range (LDR) and low frame per second (FPS) outputs. Previous works in high dynamic range (HDR) video reconstruction uses sequence of alternating exposure LDR frames as input, and align the neighbouring frames using optical flow based networks. 
However, these methods often result in motion artifacts in challenging situations. This is because, the alternate exposure frames have to be exposure matched in order to apply alignment using optical flow. Hence, over-saturation and noise in the LDR frames results in inaccurate alignment. 
To this end, we propose to align the input LDR frames using a pre-trained video frame interpolation network. This results in better alignment of LDR frames, since we circumvent the error-prone exposure matching step, and directly generate intermediate missing frames from the same exposure inputs.
Furthermore, it allows us to generate high FPS HDR videos by recursively interpolating the intermediate frames. Through this work, we propose to use video  frame interpolation for HDR video reconstruction, and present the first method to generate high FPS HDR videos. Experimental results demonstrate the efficacy of the proposed framework against optical flow based alignment methods, with an absolute improvement of 2.4 PSNR value on standard HDR video datasets \cite{10.1117/12.2040003}, \cite{Kronander2014AUF} and further benchmark our method for high FPS HDR video generation.
\end{abstract}
\begin{IEEEkeywords}
Computational Photography, High Dynamic Range, Video Processing
\end{IEEEkeywords}


%
\IEEEpeerreviewmaketitle

\section{Introduction}
Digital videos are extremely rich sources of information that can convey meaningful stories through the lens of easily accessible and portable handheld cameras devices. Two roadblocks stand in the way of achieving this full potential - low dynamic range and sub-optimal frame rates. Due to limits in memory resources of the digital sensors, standard off-the-shelf cameras can capture only a narrow spectrum of light intensity levels (dynamic range) in a natural scene and only at limited frame rate.

Low dynamic range of standard camera sensors leads to loss of pixel information in the high and low illuminance regions of a scene, resulting in overexposed or underexposed images. And low frame rate cameras are unable to capture the swift movements of a scene and lose temporal precision. High speed (FPS) HDR videos can therefore widely benefit vision systems for autonomous vehicles, movie production, hand-held cameras for recording personal moments, and much more.

Recent works have explored many variations in convolutional neural networks (CNNs) for reconstructing artifact-free HDR images from multi-exposure bracketed images \cite{LearningHDR, DBLP:conf/eccv/WuXTT18, 8579036, gong2019ahdr} and even from single exposure LDR input images \cite{EKDMU17, endoSA2017, Khan2019FHDRHI, Marcel:2020:LDRHDR}. These methods have been able to perform exceedingly well and generate high quality HDR images. The problem of HDR video reconstruction has however not been sufficiently explored, except for a few works \cite{10.1145/1201775.882270, Mangiat2010HighDR, HDRVideo, 7792731, Kalantari2019DeepHV}.  Mainly all the previous works follow a common approach of matching the exposure of neighbouring frames to perform image alignment, and merging the aligned frames to generate an HDR video. Due to the alternating exposure nature of the source LDR frames, the high exposure frames generally have oversaturated regions and the low exposure frames have high frequency noise. Therefore, even after exposure matching, images have missing regions, which results in inaccurate alignment of the LDR images. This results in motion blur in the final HDR video.

In this work we show how in the right setting, HDR reconstruction can benefit from synthesising intermediate frames using video
frame interpolation. We propose a novel approach for augmenting
dynamic range by generating intermediate alternating exposure
frames and merging them using an attention-based merge network. The pre-trained video interpolation network transfers knowledge from large scale pre-training on high FPS video datasets, and significantly improves motion compensation. 
This not only facilitates better HDR video synthesis, but can also be used to recursively increase the frame-rate of the generated HDR videos during inference. 

So far, generating high FPS HDR videos has only been possible
using event cameras \cite{Rebecq2021HighSA} \cite{Rebecq_2019_CVPR}. However, they suffer
from low spatial resolution, and are very expensive and therefore less accessible. Our approach, however, can be used to convert an alternating exposure video captured from off-the-shelf cameras into a high FPS HDR video. 

In summary, our key contributions are:

\begin{enumerate}
\item We present a novel HDR video reconstruction framework for augmenting dynamic range by interpolating LDR video frames, instead of relying on error-prone exposure matching and image registration techniques for image alignment. 
\item We present the first deep learning framework for HDR video frame interpolation, which can generate high resolution HDR videos at arbitrarily high frame-rates. 
\item We present both quantitative and qualitative experiments to validate the superiority of the proposed HDR video generation method at both standard and high frame-rates.

\end{enumerate}


\section{Related Works}
\noindent\paragraph{\textbf{HDR imaging}} HDR image generation has been extensively studied in the last few decades. Two major approaches have been adopted - 1) motion pixel rejection (\cite{article}, \cite{Jacobs2008AutomaticHR}, \cite{5693088},  \cite{10.1007/s00371-011-0653-0}) and 2) image alignment before merging (\cite{903475}, \cite{10.1007/978-3-642-33718-5_36}, \cite{10.1145/2366145.2366222, 6618998}, \cite{7301366}).
However these approaches either lose important information, when rejecting pixels, or fail to account for large motions in challenging cases, while performing image alignment. In the last few years, many deep convolutional neural network (CNN) architectures have been utilised to compose high quality HDR images. Methods like \cite{LearningHDR, DBLP:conf/eccv/WuXTT18, gong2019ahdr}, \cite{8658831}, \cite{8989959}, \cite{endoSA2017}, \cite{Niu2021HDRGANHI} have been successful in generating high quality HDR images using multi-exposure input images and have been able to outperform classical approaches. 
The abilities of these deep networks have also been extended to generate decent HDR images from single LDR images \cite{EKDMU17,Lee2018DeepRH,8457442,10.1111:cgf.13340,Khan2019FHDRHI,Marcel:2020:LDRHDR}, \cite{Liu2020SingleImageHR}, \cite{Kim2021EndtoEndDL}. Recently, an attention-based network was proposed by \cite{gong2019ahdr} to eliminate motion discrepancies among multi-exposure images with large displacements to generate ghosting-free HDR images. 


\noindent\paragraph{\textbf{HDR video generation}} Pioneer works in HDR video reconstruction, such as \cite{10.1145/1201775.882270}, \cite{Mangiat2010HighDR} and \cite{HDRVideo}, generated HDR video frames from alternating exposure input by using classical flow based approach to align neighbouring images to the reference image. \cite{7792731} proposed to avoid exact correspondence estimation and use a maxium aposteriori estimation (MAP) resulting in superior performance over flow based methods. This process, however, entails heavy computation load. Some works like \cite{liu2018switchable}, \cite{eilertsen2019single} focus on generating HDR videos from singly exposed LDR videos.
 These methods exploit single image HDR generation methods and extend them to work with videos. Lately, works like \cite{Kim2019DeepSJ}, \cite{articlejsi}, \cite{chen2021new} solve a closely related problem of SDRTV-to-HDRTV where HDR video is generated with HDR display in pixel domain unlike the standard LDR to HDR video generation in linear domain. 

Recently, \cite{Kalantari2019DeepHV} proposed a deep CNN to generate HDR videos from alternating exposure LDR sequences. They were able to outperform existing methods and reduce the processing time by several orders using a CNN based flow-based network followed by an HDR merge network. Their merge network predict the weights of input LDR images to merge.
Their approach is, however, unable to generate good results in challenging conditions with complex motion and high saturation (see Figures \ref{fig:poker2} and \ref{fig:fireworks_comparison}). Authors of \cite{Chen2021HDRVR} used the same flow and HDR merge network proposed by \cite{Kalantari2019DeepHV} and add deformable convolution module to improve the alignment.

In contrast to previous works, we propose to replace image alignment and registration based motion compensation with video frame interpolation which is used to generate alternate exposure images at each timestamp. Then, we merge the images using an attention-based merge network that synthesises HDR images from scratch instead of predicting weights of the input images. 

\section{Method}
\label{method}

\begin{figure}[!t]
    \centering
    \includegraphics[scale=.45]{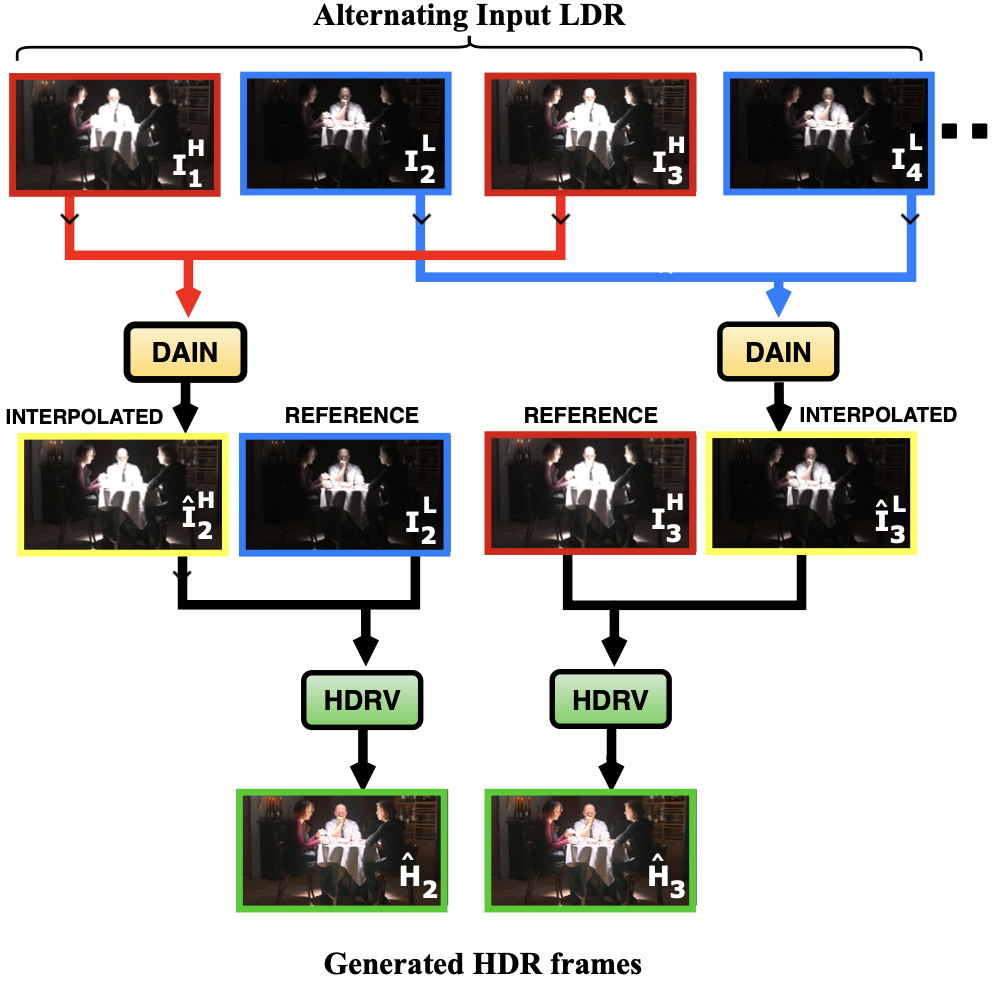}
    \caption{Overview of the proposed method}
    \label{fig:my_label1}
\end{figure}

\subsection{Overview}

 Given a video sequence of alternating exposure LDR images, $I^H_{1}$, $I^L_{2}$, $I^H_{3}$, $I^H_{4}$, .., $I^H_{n}$, our goal is to generate corresponding ground truth HDR video sequence, $H_{1}$, $H_{2}$, $H_{3}$, $H_{4}$, .., $H_{n}$. Here $I^{H}_{t}$ and $I^{L}_{t}$ denote the high and low exposure LDR images respectively at time step $T = t$. 

Having a triplet of consecutive alternating exposure images $[I^H_{t-1}$, $I^L_{t}$, $I^H_{t+1}]$, the middle one being used as the reference, we aim to generate an HDR image $\hat{H}_{t}$ that is aligned with the reference LDR frame at time $T = t$. To achieve this, we require both the high and low exposure frames at the reference timestamp. Since we already have the reference image, $I^L_{t}$ at the base exposure, we artificially generate the alternate exposed image at the same instance $t$, i.e. $\hat{I}^H_{t}$. We do so by using a video frame interpolation network to generate $\hat{I}^H_{t}$ from the similarly exposed input neighbouring frames, $I^H_{t-1}$ and $I^H_{t+1}$. Now that we have two alternately exposed images at time $T = t$,  we propose to fuse them using an attention-based HDR merge network to generate an HDR image $\hat{H}_{t}$. The end-to-end framework can be visualised in Figure \ref{fig:my_label1}.

\subsection{Video frame interpolation and HDR merge network}
For synthesising intermediate frames, we use the Depth Aware Video Interpolation (DAIN) method \cite{DAIN}. DAIN is a deep network that leverages depth information to formulate occlusion reasoning. This allows it to emphasise sampling of closer objects and facilitates robust frame interpolation. After generating an intermediate frame, the next part of the process is to merge the reference and the interpolated frame using an attention-based HDR merge network.

Attention networks for HDR reconstruction have proven to be state-of-the-art for learning attention maps to guide HDR image generation \cite{gong2019ahdr}. These learned soft attention maps are adept in dealing with misaligned LDR images and also help in picking appropriately exposed, unsaturated regions from each exposure variant. 

\begin{figure}[!t]
    \centering
    \includegraphics[width = 3.2in]{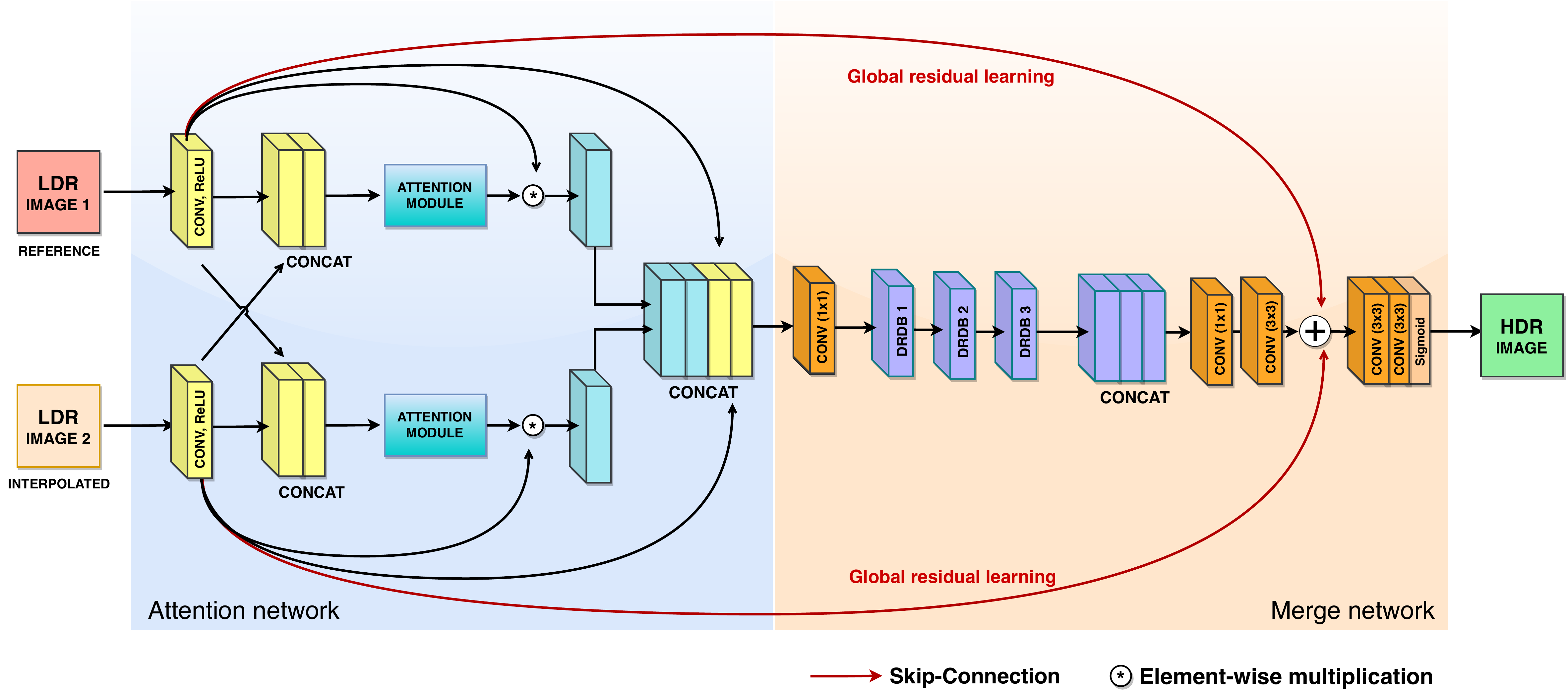}
     \caption{\textbf{Architecture of the attention-based HDR merge network (HDRV).} Given two alternate exposure LDR images, we first use the attention network to generate attention maps. These maps are then applied on the extracted features of the input LDR image. After concatenating the attention applied features and originally extracted features, these are propagated to the deep merge network.}
    \label{fig:my_label}
\end{figure}

The authors of \cite{gong2019ahdr} generate attention for high and low exposure images with respect to a reference exposure image. 
We instead do so in an alternate exposure setting. as shown in Figure \ref{fig:my_label}, we generate attention directly between the high and low exposure images. We learn two attention units for attending to high exposure and low exposure images with respect to each other, and use that information to drive HDR reconstruction in a way that reduces saturation and alignment artifacts. For more details on DAIN and attention based HDR merge network, we refer the interested readers to \cite{DAIN} and \cite{gong2019ahdr}.

\subsection{High FPS Video Generation}
In this section, we explore how the aforementioned networks can come together to generate high FPS HDR videos during inference. For example, to generate an intermediate HDR frame at time $T = 2.5$, we first generate the LDR images of the missing exposure at time $T = 2$ and $T = 3$ as shown below.
\begin{equation}
\hat{I}^H_{2} = f_{interp}(I^H_{1}, I^H_{3})\label{eq}
\end{equation}
\begin{equation}
\hat{I}^L_{3} = f_{interp}(I^L_{2}, I^L_{4})\newline
\end{equation}

Now we have both exposure images at time $T = 2$, i.e. ($\hat{I}^H_{2}$, $I^L_{2}$) and $T = 3$, i.e ($I^H_{3}$, $\hat{I}^L_{3}$). We can use similar exposure frames at both instances ($\hat{I}^H_{2}$, $I^H_{3}$ and $I^L_{2}$, $\hat{I}^L_{3}$) to generate intermediate frames of both exposures at $T = 2.5$, as shown below.
\begin{equation}
\hat{I}^H_{2.5} = f_{interp}(\hat{I}^H_{2}, I^H_{3})
\end{equation}
\begin{equation}
\hat{I}^L_{2.5} = f_{interp}(I^L_{2}, \hat{I}^L_{3})
\end{equation}
We can then merge the generated intermediate LDR frames (${I}^H_{2.5}$ and ${I}^L_{2.5}$) to obtain the final HDR image, $\hat{H}_{2.5}$ at time $T = 2.5$, as shown below.
\begin{equation}
\hat{H}_{2.5} = f_{HDR}(\hat{I}^H_{2.5}, \hat{I}^L_{2.5})\newline
\end{equation}

This procedure can be extended to generate intermediate HDR frames recursively and can thus increase the frame-rate of HDR video.

\section{Implementation details}

\subsection{Dataset}
The aforementioned network is trained on two publicly available HDR video datasets - \cite{10.1117/12.2040003} (13 scenes) and \cite{Kronander2014AUF} (8 scenes). Just like \cite{Kalantari2019DeepHV}, we reserve four scenes for the test set - CAROUSEL FIREWORKS, FISHING LONGSHOT, POKER FULLSHOT, and POKER TRAVELLING SLOW-MOTION. We prepare the dataset using the same approach described in \cite{Kalantari2019DeepHV}. From three consecutive HDR frames we generate 3 LDR images in an alternating exposure setting,
with exposures separated by one, two, and three stops. Where we choose a low exposure time randomly selected from a base high
exposure. The HDR image is delinearised using  $\gamma = 2.2$. We create randomly cropped patches from each video of size 256$\times$256. We
also apply geometric transformations like 90 degrees rotation and
flipping to augment the training data. Training data consists of about 21000 patches.

\begin{figure*}[!t]
    \centering
    \includegraphics[width=6.4in]{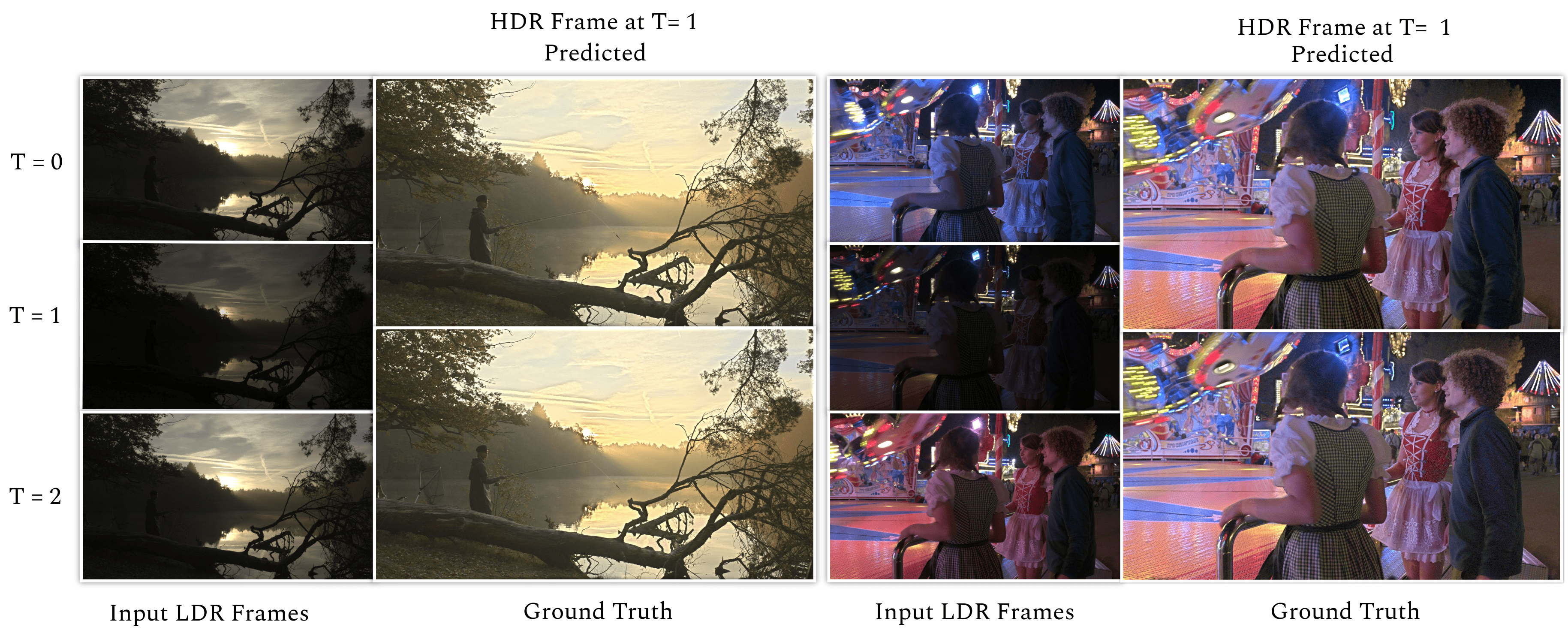}
    \caption{HDR video frames generated from the proposed approach.}
    \label{fig:fishing}
\end{figure*}

\subsection{Loss functions}
We calculate an L1 loss between the tonemapped generated HDR frame and the tonemapped ground truth HDR frame. It is imperative to tonemap the HDR images before calculating the loss because HDR loss is likely to be misrepesented due to the dominance of the highly illuminated regions in images with higher dynamic range.

As suggested by \cite{LearningHDR}, we use $\mu$ law for tonemapping because of it's differentiable nature. $\mu$ law brings the HDR data in the log domain, as shown below.
\begin{equation}
T(\hat{H}_{t}) =  \frac{log(1+\mu \hat{H}_{t})}{log(1+\mu)}\label{mu}
\end{equation}

Here, $T$ represents the tonemapping function and $\mu$ represents the amount of compression to be applied. We set $\mu$ as 5000 for our implementation.

\subsection{Training strategy}
We incorporate the following settings for training the HDR video network. For generating the intermediate LDR frames, we incorporate DAIN \cite{DAIN} as the video interpolation network, pre-trained on Vimeo90K dataset \cite{xue2019video}. We only train the HDR merge network while keeping DAIN freezed.
We use AdaMax optimizer \cite{adamax} with $\beta_{1}$ and $\beta_{2}$ as 0.9 and 0.999 with the initial learning rate set to 5$e$-4. We trained the model for 40 epochs, with a batch-size of 6. It takes about 50 hours to train the model on an Nvidia RTX 5000 GPU. After 30 epochs the learning rate is reduced by half. At inference, it takes 5.6 seconds to generate one HDR frame of size 1900x1060.

\section{Experiments}

We categorise the experiments into two parts (i) HDR video synthesis and (ii) High FPS HDR video synthesis. In the first subsection, we provide quantitative and qualitative comparisons against \cite{Kalantari2019DeepHV} for HDR video synthesis.
In the second subsection,  we evaluate our method for high FPS HDR video generation. 
We use three standard metrics for quantitative evaluation of the synthesized HDR data - PSNR, HDR-VDP-2 \cite{10.1145/2010324.1964935}, HDR-VQM \cite{10.1016/j.image.2015.04.009}. PSNR is calculated on the tonemapped HDR images using Equation \ref{mu}, while the HDR-VDP-2 and HDR-VQM scores are calculated directly on the HDR frames. For displaying, we tonemap all the HDR images using the method of \cite{Reinhard2002PhotographicTR} with modifications as suggested by \cite{10.1145/1201775.882270}.

\subsection{Evaluation of HDR video synthesis at standard
frame rates}
We evaluate our model on three seconds of the four publicly available HDR video sequences from \cite{10.1117/12.2040003}, as described in section 4. 
For quantitative evaluation, we report the results of \cite{Kalantari2019DeepHV} on the generated output
provided by the authors. For fair comparison, we choose the base high exposure of our input similar to that of \cite{Kalantari2019DeepHV} and the low exposure to be separated by three stops.

\begin{table}[H]
\centering
\begin{tabular}{@{}lccc@{}}
\toprule
                    & PSNR  & HDR-VDP-2 & HDR-VQM \\ \midrule
Kalantari\cite{Kalantari2019DeepHV} & 40.67 & 74.15     & 85.51   \\ \midrule
Ours                & \textbf{43.03} & \textbf{79.76}     & \textbf{89.58}   \\ \bottomrule
\end{tabular}
\caption{Quantitative comparison of HDR video generation at ground-truth frame-rate.
}
\label{tab1}
\end{table}

\subsubsection{Quantitative comparison}
A quantitative comparison against the method of \cite{Kalantari2019DeepHV} for HDR video reconstruction has been presented in Table \ref{tab1}. All the values are computed for each individual frame and averaged over all the frames of the four video sequences. A notable improvement over the previous method is observed across all the three metrics. 
A higher PSNR score implies a higher pixel-level accuracy, indicating that the generated HDR frames are properly aligned with the ground truth HDR frames. HDR-VDP-2 and HDR-VQM scores implies that the generated frames are visually pleasing and temporally coherent.

\begin{figure}[h]
    \centering
    \includegraphics[width=3.2in]{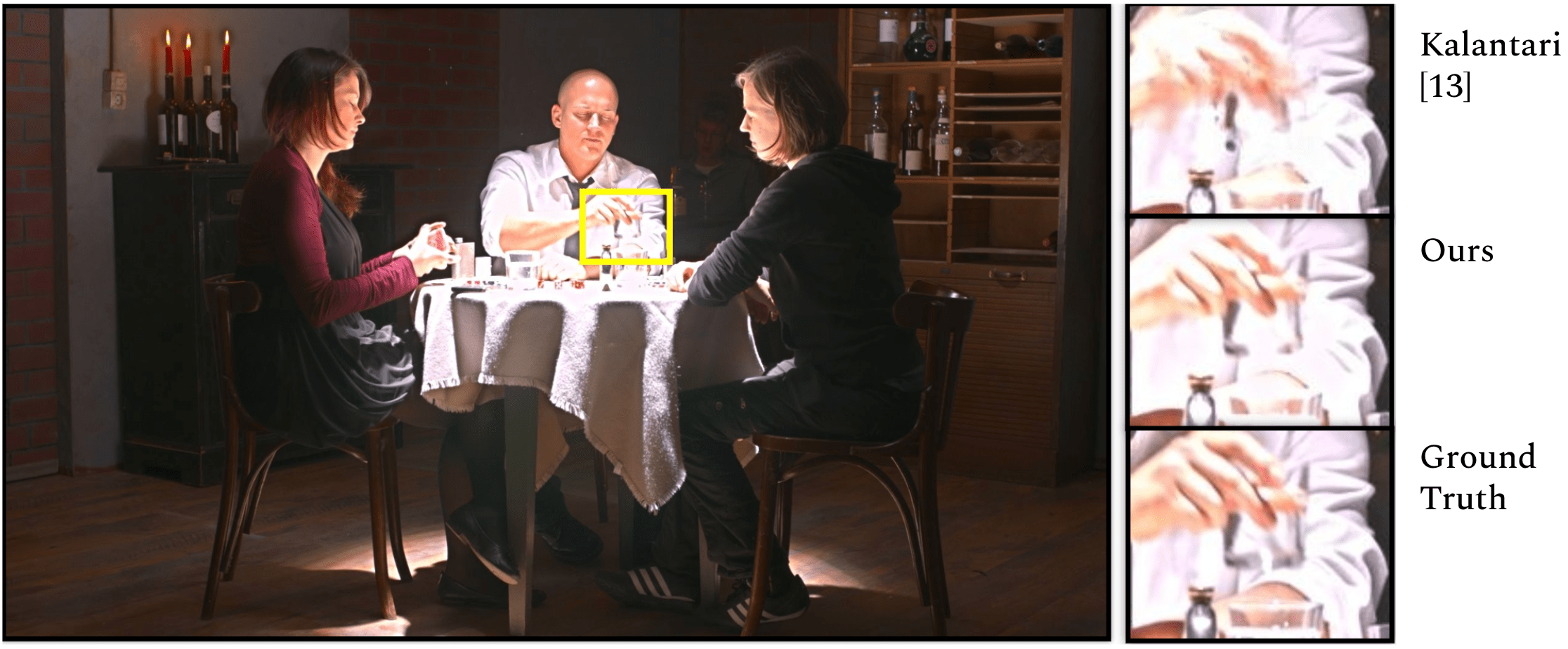}
    \caption{\textbf{Comparison for HDR video reconstruction on the POKER FULLSHOT scene.} Kalantari \emph{et al}'s method \cite{Kalantari2019DeepHV} is unable to register the fast moving hand with the glass. In contrast, our method reconstructs it accurately.}
    \label{fig:poker_comparison}
\end{figure}

\begin{figure}[h]
    \centering
    \includegraphics[width=3.2in]{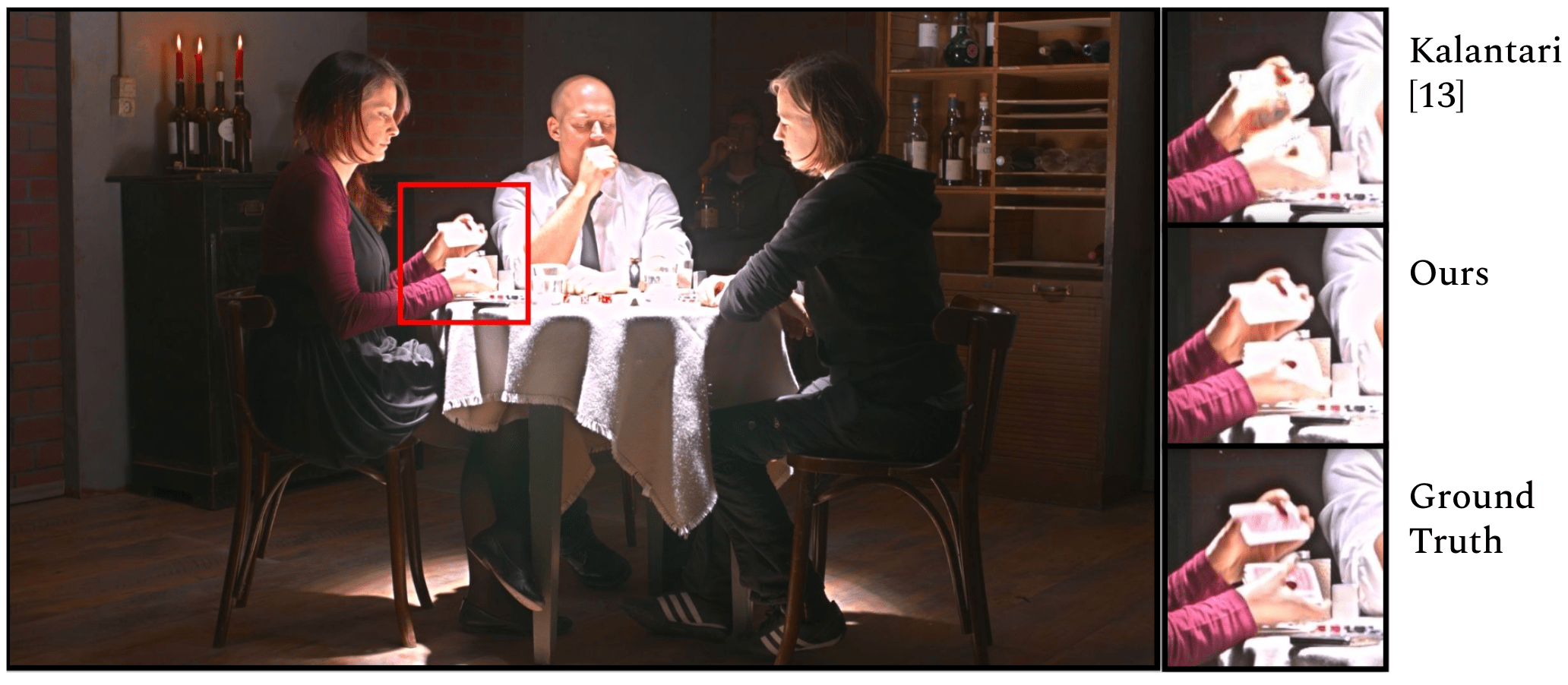}
    \caption{\textbf{Comparison for HDR video reconstrution on the POKER FULLSHOT scene.} Shape of the cards is distorted in the image generated by Kalantari \emph{et al}'s method \cite{Kalantari2019DeepHV}. On the contrary, our approach accurately reconstruct the edges of the cards.}
    \label{fig:poker2}
\end{figure}

\begin{figure}[h]
    \centering
    \includegraphics[width=3.2in]{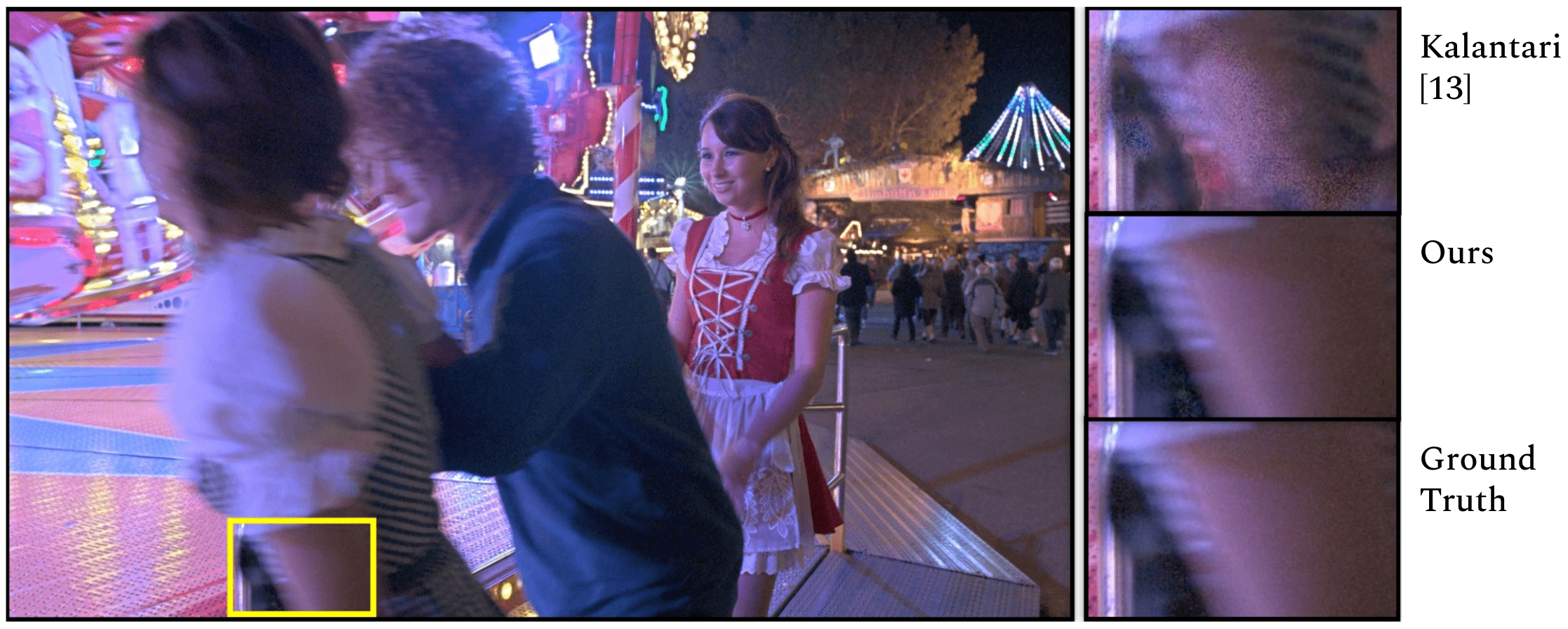}
    \caption{\textbf{Comparison for HDR video reconstruction on the CAROUSEL FIREWORKS scene}. Due to the fast motion of lady, Kalantari \emph{et al}'s method \cite{Kalantari2019DeepHV} is not able to registered the hand. On the contrary our method generates the image very close to ground truth.}
    \label{fig:fireworks_comparison}
\end{figure}

\begin{figure*}
    \centering
    \includegraphics[width=\textwidth]{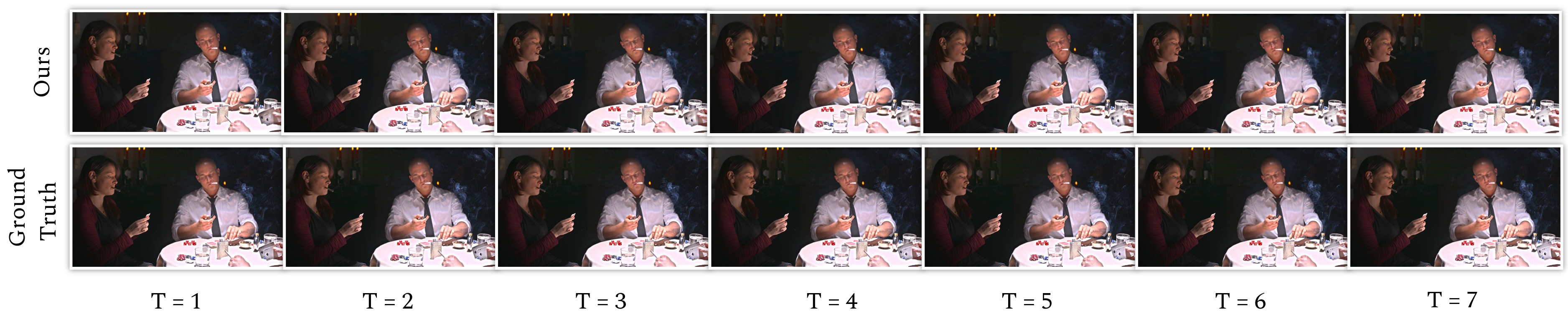}
    \caption{\textbf{Our results on the POKER TRAVELLING SLOW-MOTION scene at FPS = 8x.} We generate seven HDR intermediate frames between two input frames at $T = 0$ and $T = 8$. As can be seen, the interpolated HDR frames do not suffer from any artifacts or motion blur.}
    \label{fig:circus}
\end{figure*}

\subsubsection{Qualitative comparison}
We argue that by completely eliminating the image registration process and by simply interpolating the intermediate frames, we are able to deal with large motions. As seen in Figure \ref{fig:poker_comparison}, Kalantari \emph{et al} \cite{Kalantari2019DeepHV} is not able to register the fast moving hand of the man. Also the glass in his hand has not been synthesised accurately. 
In Figure \ref{fig:poker2}, the cards are distorted.
Similarly, in Figure \ref{fig:fireworks_comparison}, hand of the lady has not been reconstructed properly. This blur is a result of inaccuracies of the image alignment process in scenes with large motions. Our method, however, is able to preserve the boundaries and spatial details more accurately and generate sharp images, with minimum motion blur. More comparisons and qualitative results are presented in the supplementary video.

\subsection{Evaluation of High FPS HDR Video Synthesis}
To evaluate the proposed method for high FPS video generation, and validate the generality of using video frame interpolation for HDR video generation. We study the effects of three video frame interpolation alternatives when combined with our attention-based HDR merge network (HDRV) on high frame rates.

\subsubsection{Quantitative Evaluation}

In this section, we review the performance of our model in synthesising high FPS HDR video from a sequence of alternating exposure LDR frames. For this evaluation, we use the POKER TRAVELLING SLOW-MOTION scene because this is the only high FPS HDR video scene in the existing datasets. We report the PSNR, HDR-VDP-2 and, HDR-VQM scores on six seconds of this scene, for 1x, 2x, 4x, and 8x frame rates. We compare the performance of the three video interpolation networks combined with our HDRV network.
1) DAIN \cite{DAIN}, 2) Super SloMo \cite {8579036} and, 3) Modified flow network of \cite{Kalantari2019DeepHV}. The flow network proposed in \cite{Kalantari2019DeepHV}  uses three input images and generate two flow maps to align both the first and the third image to the reference image. We modify it into an interpolation network. To do so, we predict two output flow maps, and generate a visibility mask to get the final interpolated image as done in Super SloMo \cite {8579036}. We then use these pre-trained networks along with our HDR merge network.

\begin{table}[h]
\centering
\begin{tabular}{@{}lllll@{}}
\toprule
                    & FPS = 1x     &  FPS = 2x      &  FPS = 4x      &  FPS = 8x       \\ \midrule
 Kalantari\cite{Kalantari2019DeepHV}  &  44.07          &  40.41          & 37.01          &  33.54          \\ \midrule
 Super SloMo\cite{8579036}       & 45.42          &  43.33          & 40.18 &  35.57          \\ \midrule
 DAIN\cite{DAIN}              &  \textbf{46.54} &  \textbf{45.26} &  \textbf{42.22} &  \textbf{37.82} \\ \bottomrule
\end{tabular}
\caption{PSNR scores on High FPS HDR video synthesis of three video frame interpolation networks combined with our HDRV network at FPS=1x, 2x, 4x, and 8x}
\end{table}

\begin{table}[h]
\centering
\begin{tabular}{@{}lllll@{}}
\toprule
& FPS = 1x       & FPS = 2x       & FPS = 4x       & FPS = 8x       \\ \midrule
Kalantari\cite{Kalantari2019DeepHV} & 76.35          & 69.67          & 61.2           & 56.4           \\ \midrule
Super SloMo\cite{8579036}        & 81.07          & 75.56          & 68.6           & 61.25         \\ \midrule
DAIN\cite{DAIN}                & \textbf{84.27} & \textbf{80.51} & \textbf{73.59} & \textbf{65.46} \\ \bottomrule
\end{tabular}
\caption{HDR-VDP-2 scores at FPS=1x, 2x, 4x, and 8x}
\end{table}

\begin{table}[h]
\centering
\begin{tabular}{@{}lllll@{}}
\toprule
                     &  FPS = 1x    &  FPS = 2x       &  FPS = 4x       &  FPS = 8x       \\ \midrule
 Kalantari\cite{Kalantari2019DeepHV} &  84.11       &  82.51          &  80.76          &  77.89          \\ \midrule
 Super SloMo\cite{8579036}        &  84.94       &  83.88          &  82.43          &  80.09          \\ \midrule
 DAIN\cite{DAIN}                &  \textbf{86} &  \textbf{85.58} &  \textbf{84.59} &  \textbf{82.74} \\ \bottomrule
\end{tabular}
\caption{HDR-VQM scores at FPS=1x, 2x, 4x, and 8x}
\end{table}

As seen in the Tables 2, 3, and 4, we are able to generate high frame-rate HDR videos from all the three models with good accuracy. We achieve the best results on (DAIN + HDRV) model. As we move towards higher FPS, we see a drop in the accuracy. This is due to the fact that interpolating intermediate frames gets more difficult with more sparsely sampled input. However, the videos generated are temporally coherent and spatially sharp even for 8x frame rate upscaling. As seen in Figure \ref{fig:circus}, seven intermediate frames have been generated between the input frames at $T = 0$ and $T = 8$, using (DAIN+HDRV) upscaling the frame rate to 8x. The generated frames have smooth motion and are temporally coherent. This experiment validates the generality of using video interpolation for HDR video synthesis.

\begin{figure}[h]
    \centering
        \includegraphics[width=3.2in]{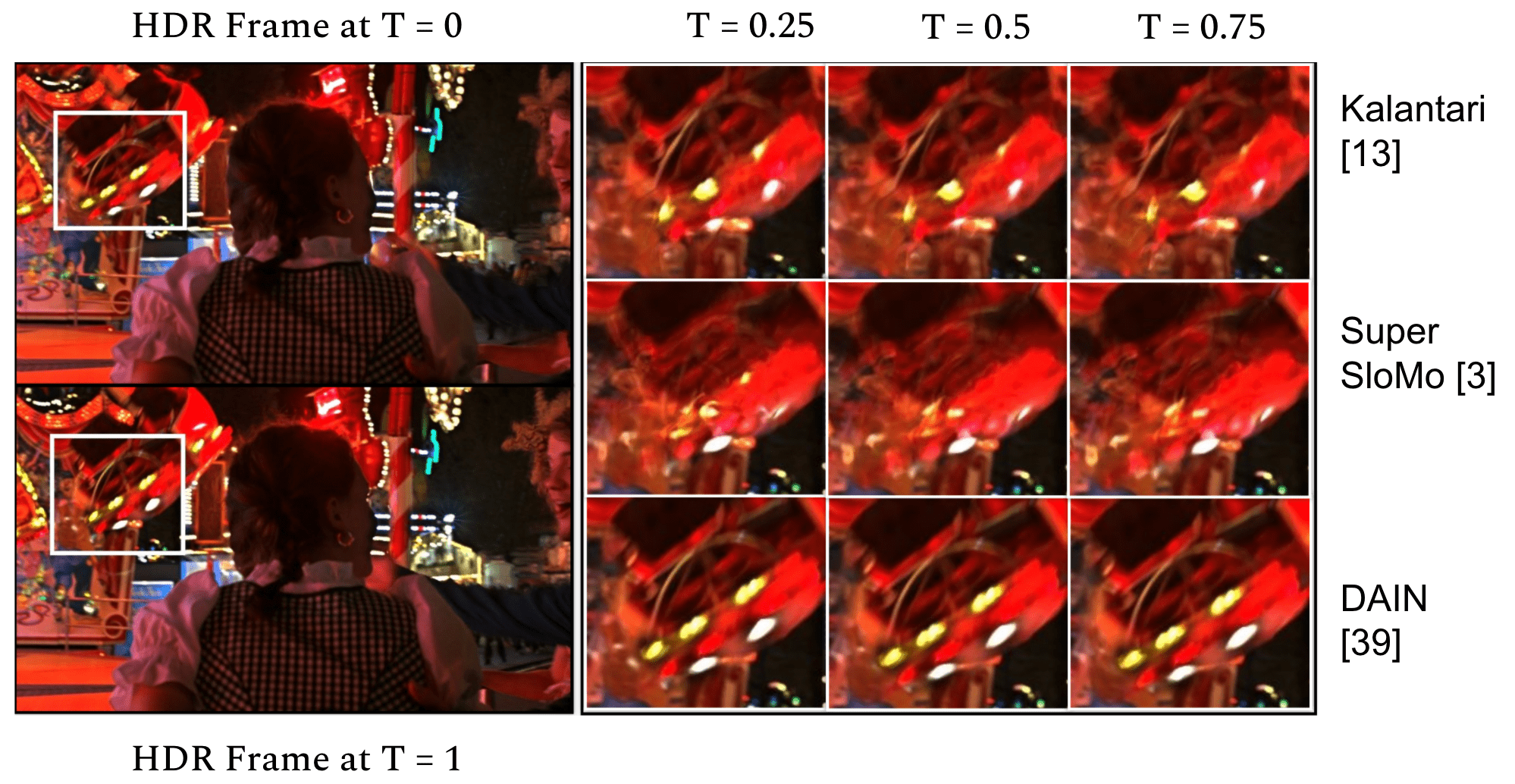}
    \caption{\textbf{Comparison for high FPS HDR video generation on the CAROUSEL FIREWORKS SCENE.} We show the results on 4x upscaled frame-rate by generating three intermediate frames between $T = 0$ and $T = 1$.}
    \label{fig:fig12}
\end{figure}
\begin{figure}[h]
    \centering
    \includegraphics[width=3.2in]{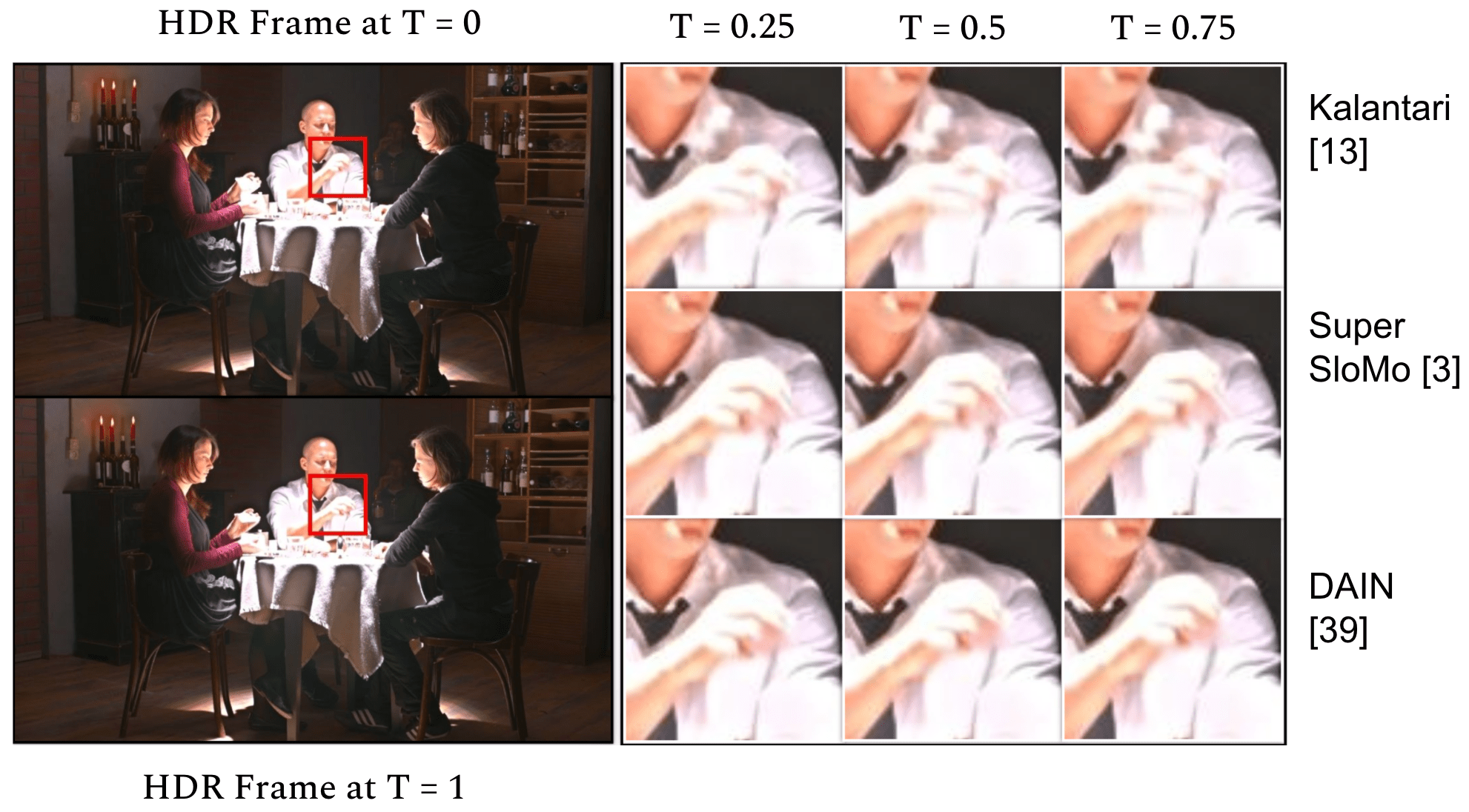}
    \caption{\textbf{Comparison for high FPS HDR video generation on the POKER FULLSHOT SCENE.}}
    \label{fig:fig13}
\end{figure}

\subsubsection{Qualitative Evaluation} 
For qualitative evaluation, we show our results on the POKER FULLSHOT and CAROUSEL FIREWORKS scenes by upscaling the frame-rate to 4x. These are low frame rate input videos with fast motion. It can be seen in Figure \ref{fig:fig12} that (DAIN+HDRV) is able to interpolate the fast moving swing of the ride with minimum motion blur. However, the interpolation networks of Kalantari \emph{et al.} \cite{Kalantari2019DeepHV} and Super SloMo \cite{8579036} produce severe artifacts in large motions. 
In Figure \ref{fig:fig13}, we can see that the hand is distorted in Kalantari \emph{et al.} approach, while Super SloMo and DAIN generate much better interpolations. This experiment suggests that the quality of high FPS HDR video synthesis heavily relies on the video frame interpolation model.
More results on HDR video frame interpolations at very high FPS upto 8x are presented in the supplementary video. 

\section{Conclusion}
We present a novel HDR video reconstruction framework from a sequence of alternating exposure LDR frames. We propose to use video frame interpolation to generate the missing exposure frames instead of relying on image registration techniques. We show that any video interpolation model pre-trained on large scale high FPS LDR video datasets can be directly used to reconstruct an HDR video. We validate its generality by performing extensive experimentation using various video interpolation methods. Moreover, we extend the proposed framework to generate high FPS HDR videos by recursively interpolating the LDR frames. 
This is the first method to generate high FPS HDR videos from off-the-shelf digital cameras.
\section{Acknowledgement}
This research was supported by SERB Core Research Grant.

\bibliographystyle{IEEEtran}
\bibliography{ICPR}
%




\end{document}